\documentclass[3p,times]{elsarticle}

\usepackage{ecrc}

\volume{00}

\firstpage{1}

\journalname{Journal}

\runauth{E. Klotins et al.}

\CopyrightLine{2011}{Published by Elsevier Ltd.}

\usepackage{amssymb}

\usepackage[figuresright]{rotating}

\usepackage{url}
\usepackage{booktabs}
\usepackage{graphicx}
\usepackage{color}
\usepackage{array}
\usepackage{hyperref}
\usepackage{ragged2e}
\usepackage{enumitem}   

\usepackage{amsmath}
\usepackage{kbordermatrix}

\usepackage[utf8]{inputenc}
\usepackage{csquotes}
\usepackage[super]{nth}
\usepackage[switch]{lineno} 
\usepackage{xspace}
\usepackage{minibox}

\usepackage{multirow}
\usepackage{tabularx}
\usepackage[numbers]{natbib}
\usepackage{rotating}
\usepackage{listings}
\usepackage{adjustbox}
\usepackage[colorinlistoftodos,prependcaption]{todonotes}

\usepackage{xcolor}

\usepackage{xparse}
\NewDocumentCommand{\codeword}{v}{%
\texttt{\textcolor{black}{#1}}%
}

\hypersetup{
    colorlinks,
    citecolor=blue,
    filecolor=blue,
    linkcolor=black,
    urlcolor=black
}

\newcounter{EKXCommentsCounter}

\newcolumntype{L}[1]{>{\raggedright\let\newline\\\arraybackslash\hspace{0pt}}m{#1}}
\newcolumntype{C}[1]{>{\centering\let\newline\\\arraybackslash\hspace{0pt}}m{#1}}

\newlist{tablelist}{enumerate}{1}
\setlist[tablelist]{leftmargin=0.15in, label*=\arabic*),itemsep=0.02in}

\begin{document}

\begin{frontmatter}



\dochead{}

\title{A Method for Identification and Ranking of Requirements Sources}


\author[1]{Eriks Klotins}
\author[1]{Veselka Boeva}
\author[1]{Krzysztof Wnuk}
\author[1]{Michael Unterkalmsteiner}
\author[1,2]{Tony Gorschek}
\author[3]{Slinger Jansen}

\address[1]{Blekinge Institute of Technology, SERL}
\address[2]{fortiss, gmbh}
\address[3]{Utrecht University}

\begin{abstract}
Requirements engineering (RE) literature acknowledges the importance of early stakeholder identification. The sources of requirements are  many  and  also  constantly  changing  as  the market  and  business  constantly change.

Identifying and consulting all stakeholders on the market is impractical; thus many companies utilize indirect data sources, e.g. documents and representatives of larger groups of stakeholders. However, companies often collect irrelevant data or develop their products based on the sub-optimal information sources that may lead to missing market opportunities.

We propose a collaborative method for identification and selection of data sources. The method consists of four steps and aims to build consensus between different perspectives in an organization. We demonstrate the use of the method with three industrial case studies. 

Our results show that the method can support the identification and selection of data sources in three ways: (1) by providing systematic steps to identify and prioritize data sources for RE, (2) by highlighting and resolving discrepancies between different perspectives in an organization, and (3) by analyzing the underlying rationale for using certain data sources.

\end{abstract}


\end{frontmatter}



\section{Introduction}

Requirements engineering (RE) aims to elicit and analyze stakeholder needs, constraints, and wishes to support the rest of software engineering activities~\cite{hofmann2001requirements,SwebookV3}. One of the first steps in requirements engineering is to identify stakeholders and the potential sources of requirements~\cite{SwebookV3}.
In dynamic markets and evolving businesses, the pool of stakeholders could be large and constantly changing~\cite{alexander2005taxonomy}. New stakeholders could emerge and existing ones become obsolete. Continuous identification and ranking of stakeholders and requirements sources is a challenge. 

Consulting with a large and diverse group of stakeholders with different interests, inconsistent needs, and varying levels of commitment is another challenge~\cite{hamka2014mobile,anwar2016stakeholders,Regnell2005} and contributes to overloaded requirements engineering~\cite{Regnell2005}. Digitalization and pervasive connectivity significantly contribute to a paradigm shift towards data-driven identification, prioritization, and management of software requirements~\cite{maalej2016toward}. This means that human requirements sources are complemented by large amounts of sensor data, for example telemetry~\cite{barik2016bones}. Therefore, in this paper we focus on \textit{requirements sources} that involve both stakeholders and data produced by sensors or software. 

The importance of early and accurate stakeholder identification is stressed by the  Software Engineering Body of Knowledge (SWEBOK~\cite{SwebookV3}), the Software Engineering Institute (SEI~\cite{ProductCMMIfor2006}), and the Rational Unified Process (RUP~\cite{KruchtenRUP}). Still, the focus is primarily on categorizing requirements sources ~\cite{ProductCMMIfor2006,ISO12207,Pressman2001,Sommerville2010,Lauesen2001} rather than exploring relationships and dynamic interactions~\cite{Sharp1999,Preiss2001}. The stakeholders are identified and listed once and considered as static entities. This view is incompatible with current trends where users drive product innovation~\cite{von2006democratizing}. Furthermore, the list of stakeholders is often composed of one person (e.g., requirements analyst) rather than consulting several roles and consensus-building.


Interactions among many stakeholders produce a continuous flow of data, such as feature ideas, feedback, problem reports, requests for specific customizations, product usage data, market analysis reports, and so on~\cite{Dahlstedt}. Overloaded with all these sources, companies often choose the path of least resistance and consult with the most accessible information sources~\cite{Klotinsc,Pacheco2012}, e.g., by inventing requirements internally~\cite{Karlsson2007} or by responding to customer-specific feature requests~\cite{Klotins2016}. However, focusing on particular customers may hinder realizing the full potential of the market and growth opportunities~\cite{Pacheco2012, Alves2006}. Finally, although some consider multiple roles (viewpoints) in requirements prioritization~\cite{Liu}, considering and balancing various data sources remains largely unexplored.

There are general purpose ranking methods, such as AHP~\cite{saaty1988analytic}, and requirements engineering specific methods, see, for example, Barbar~\cite{babar2015stakemeter} and Burnay~\cite{burnay2016stakeholders}. However, general purpose methods lack instructions for use in RE context. Existing RE specific methods are tailored for use in specific contexts (e.g. bespoke RE), and lack support for collaboration in RE and adaptation for use in new contexts.

In this paper, we present a method for identifying and ranking relevant requirements sources. The method is aimed to support a)  identification of relevant requirements sources, b) ranking of requirements sources by their relevance, c) prioritization of requirements sources in changing market needs and contextual factors, d) different perspectives.  We evaluate the proposed method in five case studies in various domains.

The rest of this paper is structured as follows. Section~\ref{sec_rw} provides background and summarizes related work in the area. Section~\ref{ch8:sec_rm} describes the research methodology steps. Section~\ref{sec_the_method} presents the method and its steps. Section 5 presents three cases and lessons learned from the validation of the method. Section 6 discusses our results.
Section~\ref{ch8:sec_conclusions} concludes the paper.


\section{Background and Related Work}\label{sec_rw}

The advent of data-driven product development has shifted the efforts of requirements identification from focusing on stakeholders (humans) as the main sources of opinions and requirements, to product usage data, sensor data, and other information generated by machines  ~\cite{groen2017crowd,jin2016understanding}. We present related work with this transformation in mind, reviewing existing methods for stakeholder identification and quantification. 

\subsection{Identifying and Analyzing Requirements Sources}

The main challenge is that not all requirements sources are equally promising or essential~\cite{Karlsson2007,Dahlstedt}.  The interactions with many stakeholders produce a continuous flow of feature ideas, feedback, problem reports, requests for specific customizations, and so on~\cite{Dahlstedt}, complemented by continuous interactions between the market, developers, and the product~\cite{alspaugh2013ongoing,groen2017crowd}. 

Moreover, connectivity and social media make it very easy to submit feedback about the software products  ~\cite{jin2016understanding}. The consequence is an evolution of requirements engineering into collaborative, data-driven, and user-centered identification, prioritization, and management of software requirements~\cite{maalej2016toward}.

Investing in documenting, analyzing, and prioritizing requirements from all data sources is impractical~\cite{Alves2006,Karlsson2007}. As a consequence, companies consult with the most accessible
sources~\cite{Klotinsc,Pacheco2012}, invent requirements internally~\cite{Karlsson2007}, or respond to customer-specific feature requests~\cite{Klotins2016,klotins2019progression}. However, for new and innovative products, candidate
data sources and their relevance may be unknown, with the consequence that companies may miss growth opportunities that market-driven products can offer~\cite{Pacheco2012, Alves2006}.

SWEBOK emphasizes the importance of considering stakeholders with diverse viewpoints~\cite{SwebookV3}. Some authors, e.g., Alexander~\cite{alexander2005taxonomy}, propose a list of stakeholder classes to aid the identification and diversification of stakeholders. However, stakeholders' knowledge, power, interest etc. should be considered as well. Diverse methods suggest stakeholder triage by their power, interest, and knowledge. Stakeholders ranking on multiple-dimensions is considered the fittest to participate in subsequent requirements elicitation activities~\cite{razali2011selecting}.

\subsection{Review of existing methods}\label{sec_rem}

A recent systematic review identified several methods for stakeholder identification and quantification~\cite{hujainah2018stakeholder}. These methods propose stakeholder high-level classes, e.g., into customers, development team, and business~\cite{babar2014stakeholder,gu2011taxonomy,Pacheco2012}, and triage them into mandatory, optional, and nice-to-have stakeholders~\cite{razali2011selecting}. However, these methods are not designed explicitly for requirements sources (both stakeholders and data), and lack support for identifying and classifying  inanimate data sources.

We performed a follow-up search based on the results from Hujainah et al.~\cite{hujainah2018stakeholder} and Bano et al.~\cite{bano2014systematic}. We used snowball sampling to identify existing stakeholder identification and classification methods. We then evaluated the found methods for suitability for identifying requirements sources and crowd-based RE~\cite{groen2017crowd,maalej2016toward}. We set forth the following evaluation criteria:

\begin{enumerate}

\item EC1: Support for identification and classification of different types of requirements sources such as individual people, large groups of people, documents, artifacts, product usage patterns, and alike~\cite{maalej2016toward,alexander2005taxonomy}. 

\item EC2: Support for collaboration between multiple analysts by capturing and illustrating different perspectives, in contrast to capturing only the consensus view from the group~\cite{aurum2003fundamental}.

\item EC3: Support for adaptation of the method for use in different contexts. The results of a method depend on how well it is suited for use in the given context~\cite{Petersen2009a}.

\end{enumerate}

We summarize our results in Table~\ref{table_meth_evaluation}. We denote a method and criteria with ``Yes'' if the method description addresses the criteria. We use ``Partial'' to convey potential method support for the criteria through an extension or adaptation. Finally, we use ''No" to denote a clear lack of support. Cases where a criterion is not applicable are denoted with ``-''.

We also extract the level of validation and evaluation presented in each study. We consider a validation relevant for  the industry if the subjects, case, validation context and scale, and evaluation method is rigorous and is representative in the industry. Validations involving, for example, down-scaled cases, students, lab experiments are considered irrelevant for industry~\cite{Ivarsson2010}.

\begin{table*}[ht]

\caption{Evaluation of existing stakeholder selection methods}  
  \label{table_meth_evaluation}
 \renewcommand{\arraystretch}{1.2}

\begin{tabular}{|L{2.5in}C{0.8in}C{0.9in}C{0.8in}L{0.7in}|}
\hline
    Method summary & Different data sources & Collaboration & Adaptation & Validation \\
    \hline
      Babar et al.~\cite{babar2014bi} analyze skill and level of interest of stakeholders & No & No & No &  None\\
     
      Razali and Anwar~\cite{razali2011selecting} consider mandatory, optional, and nice to have stakeholders and analyzes their knowledge, interest, and interpersonal skills & No & Partial & No & None\\

      Bendjenna et al.~\cite{bendjenna2012using} consider stakeholder power, legitimacy, and urgency & No & No & No & None\\

      Babar et al.~\cite{babar2015stakemeter} focus on classifying stakeholders based on their personal and professional characteristics  & No & No & No & Three relevant cases \\

      Burnay~\cite{burnay2016stakeholders} propose a taxonomy of requirements elicitation sources comprising of people, organization, artefacts, processes, and environment.   & Yes & - & No & One industrial case\\ 

      Alexander~\cite{alexander2005taxonomy} propose a model for identifying stakeholders and surrogate requirements sources  & Yes & - & Yes & None \\

      McManus~\cite{mcmanus2004stakeholder} present guiding questions to identify and classify relevant stakeholders. & Partial & No & No & None \\

      Ballejos and Montagna~\cite{ballejos2011modeling} propose a model for stakeholder representation comprising of roles, interests, influences, power, and goals & No & No & Yes & One case \\

      Lim et al.~\cite{lim2010stakenet} propose a method to identify and prioritize stakeholders using social networks analysis methods & Yes & - & Yes &  Industry survey\\
    \hline
  \end{tabular}
\end{table*}

By reviewing the existing methods, we observed several patterns. First, we observe that several methods, for example, Babar et al.~\cite{babar2014bi,babar2015stakemeter}, Razali and Anwar~\cite{razali2011selecting}, and Bendjenna et al.~\cite{bendjenna2012using}, use predefined generalized criteria for analyzing stakeholders. The proposed criteria, such as interest and communication skills, are relevant for human stakeholders and cannot be applied to artifacts. Thus, these methods are context-specific and are challenging to tailor for use in a crowd RE setting. 

Ballejos and Montagna~\cite{ballejos2011modeling} and Lim et al.~\cite{lim2010stakenet} propose methods for analyzing relationships between already known stakeholders. Professional and personal relationships are relevant only for human stakeholders and cannot be applied to inanimate objects, e.g., product usage data.

None of the reviewed methods, with the partial exception of Razali and Anwar~\cite{razali2011selecting}, support the collaboration of multiple analysts. They suggest extending the proposed method with prioritization techniques supporting collaboration, such as AHP~\cite{saaty1988analytic}.

Only some studies present industrially relevant validation or evaluation along with the proposed methodology. Working together with the practitioners and conducting at least a preliminary evaluation is important to produce industrially relevant results~\cite{Ivarsson2010}.

We conclude that none of the reviewed methods is suitable for identifying and selecting data sources for MDRE. Our work occupies this research gap.

\subsection{Decision-making scenarios}

Software engineering is a collaborative activity and requires cooperation between multiple individuals~\cite{whitehead2007collaboration}. Requirements engineering is decision-making intensive ~\cite{aurum2003fundamental}. We can extrapolate decision making into the following two scenarios ~\cite{matsatsinis2001mcda}:

\textit{The group-based approaches} aim to achieve consensus by discussion and negotiation between the group members. This approach carries the risk that the more powerful members of the group dictate the final outcome. The risk is particularly high in groups consisting of individuals from different levels of an organizational hierarchy. Negotiations and discussions are possible only in relatively small groups, e.g., agile teams. Furthermore, individual views in the group could be conflicting to the degree that a consensus is impossible without compromising the final decision or rejecting conflicting decision items. Thus, group-based approaches are feasible in relatively small and rather aligned groups.

\textit{Individual-based approaches} aim to support individuals in communicating their perspectives and aggregating individual views into a final decision. Such an approach treats each individual perspective as equal regardless of rank and negotiation power. However, individual-based approaches suffer from the neglect of interactions, exchanges of ideas, and arguments that arise from discussions in the group.

As shown by Tindale et al.~\cite{tindale2000social}, the degree to which ideas, preferences, knowledge, etc., are shared in the group determines the acceptance of the final decision and affects the quality and accuracy of the final decision. However, group discussions are not feasible in MDRE, mainly due to many stakeholders and large quantities of product-usage data. Thus, eliciting and aggregating individual perspectives is a more promising approach.

Several authors explored the applicability of  Multi-Criteria Analysis Methods for decision making, e.g. for sustainable transportation ~\cite{Broniewicz}, engineering management ~\cite{Triantaphyllou}, or emerging technologies within ICT ~\cite{5353151}. However, to the best of our knowledge no study has attempted to use multi-criteria analysis methods for software engineering and, in particular for selecting potential requirements sources.

\section{Research Methodology}\label{ch8:sec_rm}

We follow the Design Science Research Method (DSRM) by Wieringa~\cite{wieringa2014design}. DSRM describes the steps to design an artifact considering two perspectives: i) a specific real-life situation requiring a practical solution, and ii) an abstract, research view of the problem and solution. 
In the following subsections, we describe each of the six steps in detail.

\subsection{Research objectives}

In this section, we formulate the research problem with a template proposed by Wieringa~\cite{wieringa2014design}. The objective of this study is to:
  \begin{itemize}
   \item Improve the identification and ranking of requirements sources 
   \item with a systematic method
   \item ensuring transparency and building consensus between different perspectives
   \item for a given task.

  \end{itemize}

To guide our research, we formulate the following research questions:

  \begin{description}

    


    \item{RQ1:} What are the needs towards identifying and selecting the most relevant requirements sources?
    \\\textit{Rationale:} By answering this research question, we formulate specific requirements for the method.

    \item{RQ2:} How  can a method support identification and selection of the most relevant requirements sources?
    \\\textit{Rationale:} By answering this research question, we aim to propose a method design.

    \item{RQ3:} What improvements to the method are needed for its industrial use? 
    \\\textit{Rationale:} We aim to validate the method in an industrial setting.

  \end{description}

  The research questions are answered using the design science methodology~\cite{wieringa2014design}. We identify the specific challenges of selecting stakeholders and data sources and review existing methods in Section~\ref{sec_rw}. We outline specific objectives of the method in Section~\ref{sec_objectives}; the design steps are presented in Section~\ref{sec_the_method}. Objectives and steps of the validation are outlined in Section~\ref{sec_step4}, the results of the validation are presented in Section~\ref{sec_case_studies}. We discuss our results in Section~\ref{sec_discussion}.

\subsection{Step 1: Problem identification and motivation}

This research is inspired by related work on software-intensive product engineering in MDRE and known challenges in MDRE~\cite{Karlsson2007}. Several authors, e.g., Klotins et al.~\cite{Klotins2016,klotins2019progression}, Jin et al.~\cite{jin2016understanding}, and Groen et al.~\cite{groen2017crowd}, highlight the identification and selection of data sources as one of the critical practices to ensure that the product is technically and commercially successful. Identifying the right sources of requirements remains an inherent challenge in MDRE ~\cite{Karlsson2007}. 
As summarized in Table~\ref{table_meth_evaluation}, there is a shortage of methods for identifying requirements sources. In particular, there is a need to develop  methods that can support different types of data, collaborate between multiple analysts by capturing and illustrating different perspectives and offer adaptation for various contexts. Identification of appropriate requirements sources is central to understanding the context in which the product is developed and
operated~\cite{tovar2006stakeholder}.


\subsection{Step 2: Objectives of the method}\label{sec_objectives}

The main objective is to support the identification and ranking of requirements sources. We decompose the main objective into sub-goals:

\begin{enumerate}
  
  \item Objective 1: Support for the application of the method for a given task or a problem 

  \item Objective 2: Support for identification and ranking of relevant requirements sources
  
  \item Objective 3: Support for different types of data sources such as individuals people, large groups, documents, artifacts, product usage data, and alike 
  
  \item Objective 4: Support for collaboration and consensus-building among multiple analysts but also capturing the individual viewpoints and disagreements 
  
  \item Objective 5: Transparency and understandability to comprehend the method and its output.

\end{enumerate}

\subsection{Step 3: Design of the method}

The method was incrementally designed in a series of review rounds. The authors discussed design concerns at each round, evaluated different possible solutions, and updated the method accordingly. We documented rationales behind our design choices and discussed potential alternative solutions, see Section~\ref{sec_the_method}.

\subsection{Step 4: Demonstration}\label{sec_step4}

We perform a static validation of the method to evaluate its usability and collect feedback for further development of the method~\cite{Gorschek2006}. The validation consists of the following steps:

\begin{enumerate}

  \item We start by presenting the aims of our study and establishing a common vocabulary. Throughout the demonstration, we avoid posing our views, instead, we elicit participants' perspectives on data sources identification and selection, utilized practices and challenges from their experience.

  \item We ask the participants to describe a recent example where requirements sources were selected and used in crafting ideas for further product development. We ask the participants to list all the data sources considered and, with hindsight, rank them according to their contribution to the outcome. 

  The purpose of this step is to establish a baseline for comparison with the results from the method.
  
  \item We continue by introducing the method step by step and asking participants to describe the trigger, list criteria, and candidate data sources    relevant for the example. This step is done in a discussion format inquiring participants about the motivation of selecting or ignoring certain data sources or criteria. 

  The purpose of this step is to validate constructs of the method, their understandability, and practitioners ability to provide meaningful criteria and data sources.

  \item We ask participants to provide scores for criteria, evaluations to data sources and run the calculations to arrive at the final ranking. We use a spreadsheet for data collection.

  The purpose of this step is to arrive at a ranking of data sources based on the method for comparison with the baseline.
  
  \item Reflections on the results. In the end, we ask participants to reflect on the method and its consequences, compare it with the initial ranking, and provide their perspective on the differences. 

\end{enumerate}
We conduct three case studies and describe the results in Section~\ref{sec_case_studies}. Demonstrations are handled by the first and second authors with the help of presentation materials and a semi-structured interview guide. Multiple practitioners were involved in each demonstration.

\subsection{Steps 5-6: Evaluation and communication}
Further evaluation and communication of the method are planned. We aim to follow the technology transfer model by Gorschek et al.~\cite{Gorschek2006}.

\subsection{Threats to validity}

We follow the guidelines by Runeson et al.~\cite{Runeson2012} and discuss the four perspectives of validity threats.

\subsubsection{Construct validity} Construct validity is concerned with establishing appropriate
measures to observe the intended concepts. 
Key concepts of the method presented in this paper originate
from related empirical work on innovative software-intensive product management
and related work in stakeholder identification in requirements engineering and crowd RE. The evidence we accumulated in Table \ref{table_meth_evaluation} supports building our arguments of the need for a method that is flexible (can be applied to different problems) and supports collection opinions from various roles and building consensus. Discussion with participants at the case companies also supported the formulations and relationships of the concepts. 

We further strengthen the construct validity during the case study, we asked the participants to discuss their views on practices and challenges associated with stakeholder and data sources selection.

\subsubsection{Internal validity} Internal validity is concerned with uncontrolled factors affecting causal relationships between concepts. The method infers that criteria are the only 
yardstick to rank requirements sources. There could be other influences on the ranking that do not fall under our definition of criteria. 
To minimize this threat, the method needs to be further validated and operationalized, as well as more work is required towards understanding decision-making factors in stakeholder selection.

Our participation may influence the review and responses of the participants.  Knowing the objectives of our study, practitioners may inflate the challenges of data sources selection. Furthermore, participants may be reluctant to reveal their real opinions about the method because the authors were present. To mitigate this threat, we avoid posing specific views, instead, we introduce concepts neutrally and ask practitioners to reflect on their experience. 

A potential limitation of the method is the need to avoid ties between the criteria. Having dependencies between criteria would impair rankings and the final results. We plan to explore different strategies for removing this limitation with further work.

\subsubsection{External validity} External validity is concerned with the extent to which the results can be
generalized outside the studied cases and remains the main limitation of our
work. We cannot claim that the studied cases are representative of all companies. Thus, the fitness of the method needs to be further validated. Therefore, we focus on analytical generalization, rather than statistical generalization \cite{Flyvbjerg2006}. The diversity of the three case studies further strengthen external validity. 

\subsubsection{Reliability} Reliability concerns the degree of repeatability of the study. 
To support traceability and transparency, we described  the rationale and objective for each step of the research method and the case study. We also provide raw data and our calculations as supplemental material online
\footnote{Jupyter notebook with all data and code: \url{https://github.com/EriksKlotins/data_sources_method}}

However, the method embodies ideas, knowledge, and interpretations that remain to some extent subjective.

\section{The method}\label{sec_the_method}



\begin{figure*}[!ht]
\centering
\includegraphics[width=\textwidth]{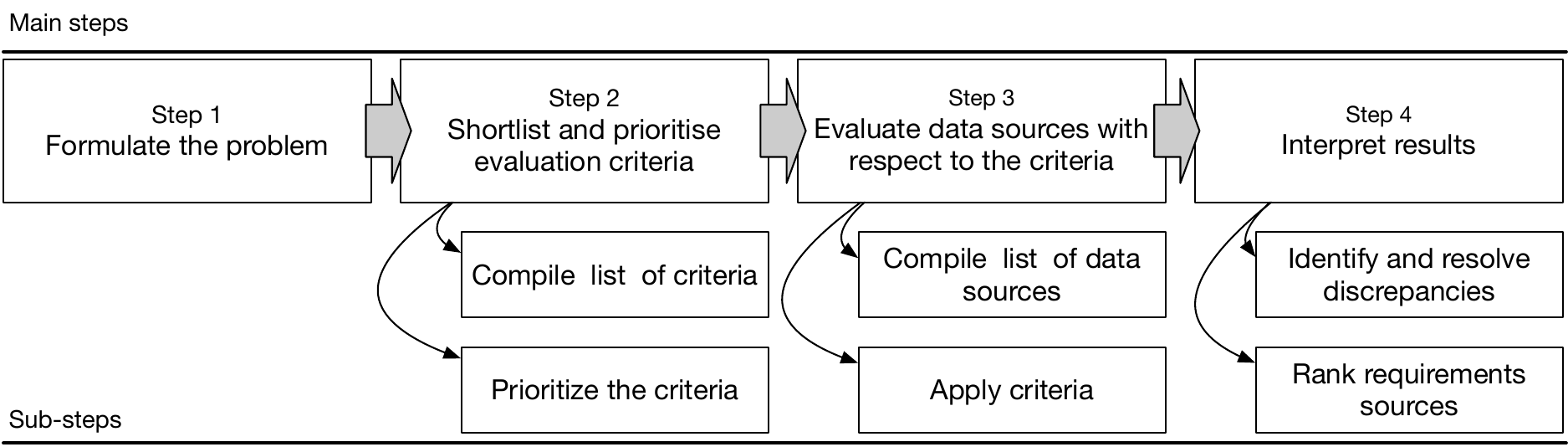}
\caption{Overview of the method. The method consists of 4 main steps with several sub-steps}
\label{fig_method}
\end{figure*}

Our method combines multiple analysts' input and constructs a ranked list of the requirements sources to analyze and consult in a given decision-making situation.  The requirements sources are rated based on their attributes, which are selected and estimated by the analysts. The method consists of four main steps, see Fig~\ref{fig_method}.

In {\it  Step 1}, the analysts identify a task and formulate a problem statement. The problem statement implies the need to decide and consult multiple stakeholders and requirements sources.
In {\it Step 2}, the analysts list relevant criteria for the task (problem) for which identification and ranking of requirements sources is needed. 
In {\it Step 3}, the analysts evaluate the requirements sources according to the identified criteria. 
In {\it Step 4}, the analysts analyze and interpret the results.

\subsection{Preconditions}

We start the method design by identifying and outlining the core concepts described below.

\textit{A trigger} is circumstances leading up to the formulation of the task or problem statement. A trigger could be, for example, the identification of a new market opportunity, a customer request for a specific feature, market pressure, or a technological shift. We believe that the dynamic nature of software business continuously generates triggers.

\textit{A problem statement} is a brief description of a non-trivial issue to be
addressed or a situation to be improved. A problem statement describes the gap
between the desired situation and the current situation~\cite{annamalai2013importance}. A problem statement should be
quantifiable, e.g., the number of lost customers, the amount of potential revenue, or saved resources.

\textit{Requirements sources} may have different forms, for example, individual stakeholders (such as key customers, managers, or product engineers), the crowd~\cite{groen2017crowd} consisting of, e.g., potential customers, opinion leaders, and users of competitor products, other organizations (such as competitors, partners, or suppliers), analytics (such as product usage data, analysis of customer data, or results from market analysis), industry standards, and applicable laws and regulations. Different requirements sources may offer different types of information and different perspectives on the decision.



\textit{Criteria} are principles of how requirements sources are selected and compared. 
The criteria depend on the task or problem to be addressed. 
For example, an internal engineer is more accessible than a customer. In turn, customers have
more profound knowledge about their actual needs than engineers but may have 
little insight into underlying technology constraints. Another example is the
constructs used by Mitchell et al.~\cite{MitchellStakeholderTheory} in his  stakeholder identification theory: power, legitimacy, urgency, and salience.  



\textit{An analyst} is an individual who collaborates with other analysts on selecting the requirements sources and on the use of the method. 
We use the term  analyst to refer to a range of people that could be involved in product decisions, such as software engineers, product managers, financial officers, marketing, sales representatives, and alike.
The analysts are involved in the decision making and are the primary stakeholders, that is, the analysts are the decision-makers. 



\subsection{Step 1: Define the problem statement}

The first step is to formulate the problem to be addressed. The problem formulation should contain the following information:

\begin{enumerate}
  \item Description of the current situation. 
  
  \item Characterization of the desired situation.
  
  \item Quantification of the difference between current and desired situation, e.g. improve $X$ by $Y$.
  
  \item Preliminary candidate solutions (high level ideas) to the problem at hand.  The candidate solutions are to be evaluated and refined by the input from selected stakeholders. 
  
  For example, a candidate solution could be to expand the offering with a new feature. Our method aids requirements analysts reach a consensus what stakeholders and other requirements sources are the most relevant for requirements elicitation.
\end{enumerate}

\subsection{Step 2: Shortlist and prioritize criteria}

The second step is to identify and prioritize the criteria for comparing potential requirements sources. This step has three sub-steps: 1) compile the initial list of criteria, 2) shortlist the relevant criteria by voting, and 3) prioritize (weight) the criteria.

\subsubsection{Compile a list of criteria}
The analysts put together an initial list of criteria that will be used to compare different requirements sources. 
The aim is to identify all relevant characteristics of an ideal data source to resolve the problem defined in Step~1. The proper criteria are domain and problem-specific and can be identified through, e.g., discussions and brainstorming. 

To facilitate the identification of relevant criteria, we adapt a taxonomy of requirements criteria proposed by Riegel et al.~\cite{riegel2015systematic}, see Table~\ref{table_criteria}. We propose to use the table's high-level criteria as a starting point for the initial analysis and brainstorming to identify additional criteria.

\begin{table*}[ht]
  \renewcommand{\arraystretch}{1.2}
  \caption{Example criteria of requirements to bootstrap data source criteria}
  \label{table_criteria}
  
  \begin{tabular}{|L{1.3in}L{1.5in}L{3.1in}|}
    \hline
    Category & High-level criteria & Examples of detailed criteria \\
    \hline
       
      Benefits in terms of knowledge  & Level, type of knowledge & Domain, customer needs, and wishes, product technologies or features, business processes, laws, regulations, standards\\
      
      Benefits in terms of experience & The amount, type of experience & The product, similar products, specific market segments, domain, product technologies \\

      Benefits in terms of financial value & Revenue, potential revenue, profit & Customer lifetime value, price per purchase \\

      Costs       & Associated costs, indirect costs            & Access, establishing access, maintaining access  \\

      Penalties  & Opportunity cost & Business, market, technical, financial, reputation, the dissatisfaction of other stakeholders \\
      
      Risks   & Generic risk & Human errors, technical risks, implementation risks, volatility, business risks, time, budget, project scope, dependencies, reputation, legitimacy  \\

      Temporal context & Timing & Lead time, time to access, timeliness of data, frequency of access \\

      Suitability for use & Ease of use & Ease of access, mutual trust, understandability, granularity, analyzability, accuracy, overhead  \\

      Behavioral & Suitability for collaboration  & Availability, interest to contribute, commitment, volatility, trustworthiness, willingness to experiment, capacity volunteer resources for collaboration, power, leverage \\

      

     \hline
  \end{tabular}
\end{table*}

To avoid inconsistencies in further prioritization and estimation, the analysts need to avoid ties between the criteria. For example, if criteria such as knowledge and domain knowledge appear together, analysts may have difficulty  assigning consistent scores~\cite{vigna2015weighted}. A counter-strategy could be to identify potential ties early and break down higher-level criteria into more fine-grained criteria, thus eliminating the ties, e.g., separate knowledge into technology expertise and market expertise.

\subsubsection{Shortlist of relevant criteria}

After the initial list of criteria is compiled, the analysts vote to shortlist the most relevant and
exclude less relevant criteria, thus reducing the effort in the next steps.
Based on the votes and a cutoff threshold, relevant criteria are selected for further consideration.

Each analyst lists all criteria and assigns a binary vote (relevant/not relevant) to each criterion.
The ballots produced by the different analysts on each criterion are summed. Criteria with votes above a given threshold are included for further consideration. For example, a majority vote can be used to decide which criteria to prioritize, i.e., we can apply a cutoff threshold of $k/2$, where $k$ is the number of analysts. 

Shortlisting may be relevant only if more than two analysts are involved or there is a strong disagreement between the analysts. For example, if a maximum of two analysts are involved, then a cutoff threshold of $k/2$ is 1. In this context, both analysts' criteria should be considered, or it may be decided to consider all criteria proposed by the analysts. 

Here and in further subsections, we exemplify the application of the method by describing analysts' choices and inputs from Case I. Analysts from the Case I selected knowledge of the product,  accessibility, trust, value per purchase, lifetime value, and the capacity to contribute as relevant criteria. Full description of the Case I is provied in Section~\ref{sec_case_study_1}.








\subsubsection{Prioritize the criteria}\label{sec_criteria_prio}

The purpose of this step is to prioritize the criteria by their relevance to the problem under a decision. Each analyst assigns a score to each criterion, assessing its relevance (importance). We propose to use an ordinal scale from 0 to 5, i.e., all scores in $\{0, 1, 2, 3, 4, 5\}$, where 0 indicates no relevance at all, 5 indicates the highest relevance, and the numbers 1, 2, 3, 4 represent intermediate values. Table~\ref{table_crit_meaning}  provides the semantic meaning of each score. 

 \begin{table}[!h]
  \renewcommand{\arraystretch}{1.1}
  \caption{Semantic meaning of the measurement scale}
  \label{table_crit_meaning}
  \centering
  
  \begin{tabular}{C{0.4in}L{3in}}
    \hline
    Score & Meaning \\
    \hline
    \renewcommand{\arraystretch}{1.1}
0 & Not relevant at all (to the given problem) \\
1 & Marginally relevant \\
2 & Somewhat relevant  \\
3 & Moderately relevant \\
4 & Very relevant \\
5 & Most relevant \\
\hline

  \end{tabular}
\end{table}







The produced individual and averaged weights can be studied and compared to better understand and interpret the
problem under consideration. It could be useful to analyze whether there is a significant discrepancy in
analysts' opinions about the criteria importance. Analyzing discrepancies can help to improve transparency, quality, and acceptance of the results~\cite{tindale2000social}.

Analysts in the Case I produced the following scores, see Table~\ref{table_crit_scores_1}.

\begin{table*}[h]
  \renewcommand{\arraystretch}{1.2}
  \caption{Criteria scores from Case I}
  \label{table_crit_scores_1}
  \centering
  \begin{tabular}{|L{1.5in}L{1in}L{1in}|}
    \hline
    Criteria & Analyst 1 & Analyst 2 \\
    \hline

Knowledge of the product &	3 &	4 \\
Accessibility &	2	& 5\\
Trust&	5	&3\\
Value per purchase &	3&	4\\
Life-time value	&4&	4\\
Capacity to contribute&	2&	3\\
\hline

\end{tabular}
\end{table*}

\subsection{Step 3: Evaluate requirements sources in regards to the criteria}
In the third step, the analysts compile a list of potential data sources and use the selected criteria to evaluate each requirements source. This step has three sub-steps: i) compile a list of requirements sources,  ii) apply criteria, and iii) calculate the result.

\subsubsection{Compile a list of requirements sources}
The analysts put together a list of potential requirements sources. The list can be created by writing down already known requirements sources, brainstorming, or a combination of both. The aim is to identify a diverse set of potentially informative requirements sources for further consideration. 

The requirements sources are usually domain-specific. However, we propose to use the high-level sources given in Table~\ref{table_sources} as a starting point for the initial analysis and brainstorming. Note that stakeholders can be accessed directly, e.g., by consulting with engineers, and indirectly, e.g., by analyzing user behavior through product usage patterns.


\begin{table*}[ht]
  \renewcommand{\arraystretch}{1.2}
  \caption{Example stakeholders and other requirements sources}
  \label{table_sources}
  \centering
  \begin{tabular}{|L{1in}L{1in}L{2.3in}|}
    \hline
    Category & High-level examples & Detailed examples \\
    \hline
      Internal stakeholders & Engineers, Product managers, business stakeholders & Product engineers, architects, customer service representatives, sales representatives, managers, company executives \\

      External stakeholders & Users & Premium customers, freemium customers, prospects, end-users, partners, competitors, suppliers, lawmakers, regulators \\

      Analytics & Product usage data & Telemetry data, user data, user behavior analysis \\

      Reports & Market research & Market analysis, public surveys, trends, analysis of similar products \\

      Environment & Domain knowledge & Domain experts, Technology standards, laws, regulations, industry conventions, opinion leaders \\

    \hline
  \end{tabular}
\end{table*}

\subsubsection{Quantify the requirements sources (apply criteria)}

The analysts apply the selected criteria to evaluate each requirements  source. Each analyst produces a table where each row maps to a requirements source, and each column to a criterion. The cells contain the analysts' score quantifying the requirements source in regards to the criterion.

We suggest using a scale from 0 to 5, where 0 denotes the lowest and 5 the highest score. Higher scores are awarded to more favorable evaluations, e.g., lower cost and higher accuracy. In Table~\ref{table_ds_meaning}, we provide the semantic meaning of each score. Note that we consider only favorable (positive) scores. 

 \begin{table}[!h]
  \renewcommand{\arraystretch}{1.1}
  \caption{Semantic meaning of the measurement scale}
  \label{table_ds_meaning}
  \centering
  
  \begin{tabular}{C{0.4in}L{3.8in}}
    \hline
    Score & Meaning \\
    \hline
    \renewcommand{\arraystretch}{1.1}
0 & The requirements source does not favorably contribute to criteria at all  \\
1 & The requirements source provide a marginal favorable contribution \\
2 & The requirements source provide a somewhat favorable contribution to the criteria  \\
3 & The requirements source provide a moderately favorable contribution to the criteria  \\
4 & The requirements source provide a favorable contribution to the criteria  \\
5 & The requirements source most favorably contribute to a criteria \\
\hline

  \end{tabular}
\end{table}

  


\subsubsection{Calculate the result}

Up to this point, each analyst has produced scores quantifying the relevance of each criterion and a table evaluating each requirements source in terms of each criterion. The resulting scores for each requirements source are calculated by normalizing analysts' scores, calculating the dot-product between data sources evaluations and criteria evaluations, and averaging the scores between multiple analysts. The averaged scores are scaled between 0--1. Python code performing the calculations is provided in Listing~\ref{method_code}. The complete code used to to produce results of the case analysis is accessible online\footnote{Jupyter notebook with data and calculations: \url{https://github.com/EriksKlotins/data_sources_method}}. The code returns intermediate results as well to enable transparency and analysis of potential disagreements between analysts.

\begin{figure}
\lstset{language=Python}
\lstset{frame=lines}
\lstset{caption={Implementation of the method calculations}}
\lstset{label={method_code}}
\lstset{basicstyle=\footnotesize}


\begin{lstlisting}
import numpy as np

# cs - criteria evaluation scores
# ds - datasource evaluation scores
# im_results = False - return intermediate results
def rank_data_sources(cs, ds, im_results = False):    
    
    result = {'csn': [], # criteria scores normalized
              'dsn': [], # data source scores normalized
              'ans': [], # resulting scores for each analyst
              'avs': []} # averaged scores
    
    # Normalize criteria scores
    result['csn'] = cs / np.sum(cs, axis=0)
    
    # for each scores matrix (i.e. Analyst)
    for i, scores in enumerate(ds):
        
        # Normalize data sources scores
        dsn = scores / np.sum(scores, axis=0)
        result['dsn'].append(dsn)
        
        # Dot-product between datasource scores 
        # and criteria scores         
        result['ans'].append(dsn @ [row[i] for row in result['csn']])
          
    n = len(result['ans'])    # number of analysts  
    
    avs = np.sum(result['ans'], axis=0) / n   # Averaged scores
    
    result['avs'] = avs / np.max(avs) # scale     
    
    return result if im_results else result['avs']
\end{lstlisting}
\end{figure}

\subsection{Step 4: Interpret results}

Interpretation of results consists of two steps: i) Identify and resolve discrepancies between analyst perspectives, ii) Present the ranked requirements sources for decision makers.

\subsubsection{Identify and resolve discrepancies between perspectives}

Breaking down the results and analyzing how different analysts have estimated criteria and requirements sources can help to spot  discrepancies. The analysis of such discrepancies can be useful for improved problem understanding, removing ambiguity, and improving results of the method. We propose to visualize the resulting scores for each analyst, i.e., \codeword{result['ans']} from Listing~\ref{method_code}. The values show how different analysts have evaluated each data source according to each criterion. Significant differences between evaluations suggest a discrepancy.

It is important to identify the cause of any discrepancies to improve and strengthen the results of the method. We recognize the following causes for discrepancies:

\begin{enumerate}
\item \textit{Mistakes.} A discrepancy can be caused by a wrongly entered number, miscalculation, or issue in a tool, etc. Such discrepancies can be easily removed by remedying the cause.

\item \textit{Misunderstanding.} Different analysts may have different interpretations of certain criteria or requirements sources. Misunderstandings can be mitigated by initiating discussions and conducting an analysis of the analysts' estimates, and making corrections if applicable. 

\item \textit{Different perspective.} Discrepancies could also be caused by different and multimodal perspectives on the problem at hand. Genuine different perspectives are handled by the method.

\end{enumerate}

Discrepancies caused by misunderstanding, misinterpretation, and perspectives modalities can be addressed by breaking down higher-level criteria and requirements sources into more specific terms. For example, a product could consist of many
technologies. Thus a criterion such as knowledge of product technology could be too broad to be used for ranking data sources. Breaking down such high-level criteria into more specific can help arrive at more consistent evaluations.

\subsubsection{Present the ranked data sources for decision makers}

The weights in the \codeword{result['avs']} indicate the overall relevance of each data source based on the evaluations of all the involved analysts. That is, higher-ranking data sources are more relevant to the problem under consideration and should be consulted with higher priority, e.g., see Fig~\ref{fig_case1beforeafter} and Fig~\ref{fig_case2beforeafter}. We suggest interpreting the results in parallel with analyzing discrepancies and exploring analysts' views on each data source and criterion. Thus, it benefits both the aggregated individual-based approach and the exchange of ideas within a group~\cite{matsatsinis2001mcda}.

\section{Case studies}\label{sec_case_studies}
In this section, we describe the demonstration and application of the method in three industrial cases and practitioners' reflections on each component of the method, lessons learned from each step, and the collected data. The three companies were selected from our network based on the following criteria: 1) they develop software-intensive product or services, 2) they undergoing significant business transformations, 3) they get the impression that the current set of requirement sources is not the most optimal for answering the challenges that the significant business transformation brings. All three companies volunteer to participated in this research receiving no compensation. 

The raw data obtained from the cases and code used to analyze the data is available online\footnote{Data and code used in the calculations:~\url{https://github.com/EriksKlotins/data_sources_method}}.

\subsection{Case I - Supply Chain Digitalization}\label{sec_case_study_1}

Stockfiller AB offers a software-intensive service digitalizing the supply chain between sellers (farmers, food producers, and food importers) and buyers (restaurants and grocery stores) of food products in Sweden. The service's main value is a more efficient and transparent supply chain that historically is based on personal contacts and manual processes. The service's monetization model is based on the volume of goods traded on the platform. The service manages about 10 000 sellers and buyers. The company has been in the market for five years. The company currently works on securing the local market before expanding to nearby regions.

Two managers of the company, chief of operations, and head of sales, participated in a workshop. Since the beginning, they have been with the company and have quite extensive knowledge about the product, market, potential requirements sources, and stakeholders.
The workshop follows the steps described in Section~\ref{sec_step4}. 
We start by eliciting practitioners' experience with selecting requirements sources and a specific example that can be used as an input to the method.



\textit{Reflections on the challenge:} The participants reflected that the quality of elicited ideas from potential customers is  low due to a lack of understanding about the service and potentially sub-optimal processes at the customer side. As a consequence, such feature requests had little value to other customers.  Moreover, customers are often incapable of articulating their actual needs. Thus demonstrating a prototype and eliciting feedback  works much better than interviews. 

The company uses knowledgeable and experienced customers as a filter to vet new feature ideas, both internal and provided by other customers. The respondents also identified product vision (their view of how the supply chain should be organized) as a critical tool to shape the product and gauge customer requests for new features. 


\textit{Triggers:} The company discovered that a substantial amount of sellers have poor sales performance on the service. Lagging sellers present missed revenue since the monetization model is based on the traded volume. Principals of the company had proposed to explore the matter.

\textit{Analysts: } The principals of the company are the main product decision-makers. Chief of operations, head of product, and director of sales regularly meet and discuss opportunities for new product features, among other concerns. In the demonstration, only the chief of operations and the director of sales were present.

\textit{Problem statement: } The principals set forth an objective to support sellers by providing insights on their sales performance. 
Initially, there were no insights for sellers to assess their performance and spot opportunities for improvement. As a consequence, the company was losing potential revenue.
The wanted scenario would be that the service provides useful information for sellers to monitor their performance and suggests adjustments to boost their trade volumes.




\textit{Criteria: } The participants immediately identified good relationships and knowledge as the primary criteria to select stakeholders for collaboration. The participants also identified the product's knowledge and experience, accessibility, value per purchase, life-time value, mutual trust, openness, capacity, and eagerness to contribute, e.g. by piloting experimental features, as relevant criteria. 


Only two analysts participated in the workshop, and disagreements on what are the relevant criteria were resolved in the discussions; thus, explicit shortlisting of criteria was not necessary.

  




Further discussions revealed that some criteria could be broken down to be more specific and suit both stakeholders and data sources, e.g., accessibility could be interpreted as the ease of access (for meetings, observations, or reading), and understandability (of the actual needs and descriptions).

\begin{figure*}[h]
      \centering
      \includegraphics[width=\textwidth]{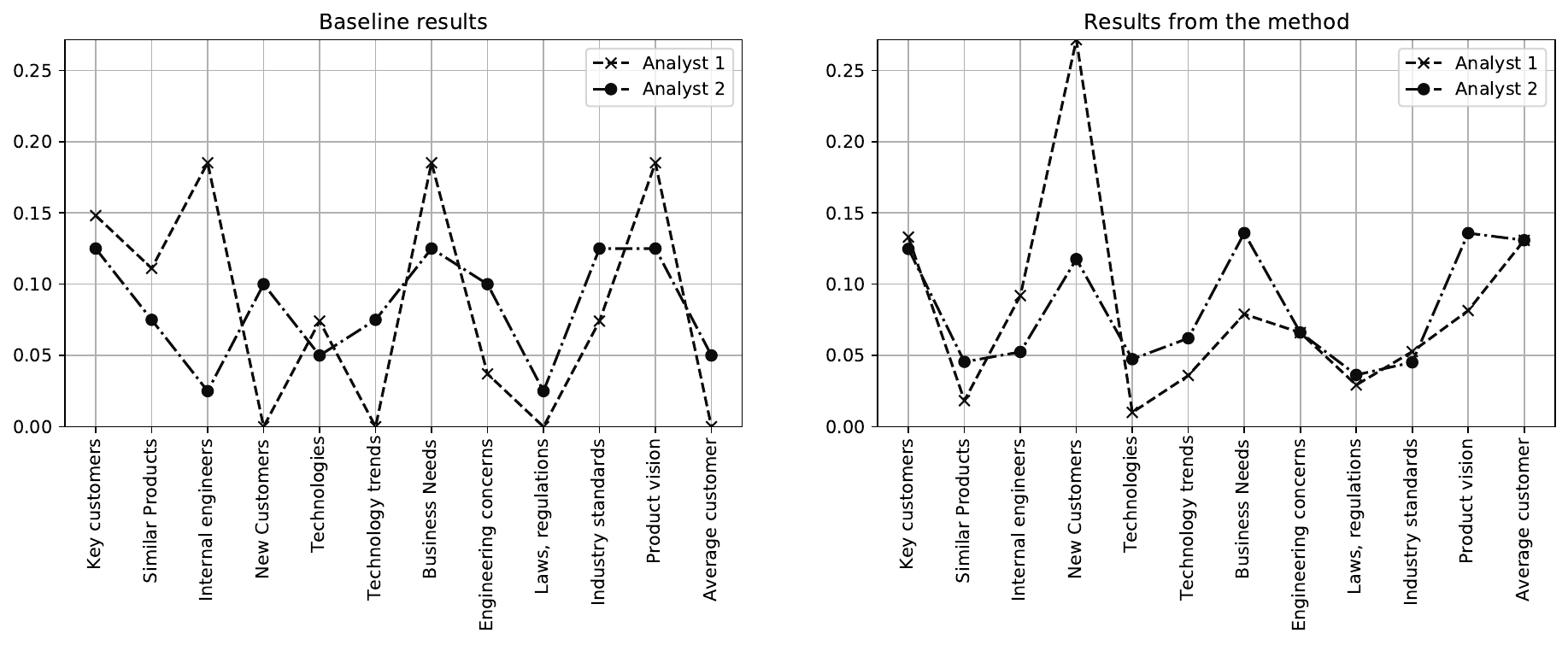}
      \caption{Results from the Case I. 
      Figure on the left shows baseline estimates, i.e. respondent estimates without the method. Figure on the right shows results from applying the method. 
      Y-axis denotes the relative importance of data sources. 
      }
      \label{fig_case1beforeafter}
\end{figure*}

\textit{Requirements sources: } The participants acknowledged using multiple inputs in their decisions. They listed experienced and knowledgeable customers (key customers), similar products, internal engineers, new customers, technologies (used design templates, patterns, frameworks), technology trends (such as new available technologies), business needs, engineering concerns, laws, and industry standards as relevant data sources. In Fig.~\ref{fig_case1beforeafter} (left side) we show the original baseline estimates on data source relevance by both analysts.

\textit{Application of the method: } The participants had no difficulty understanding the rationale behind assigning the scores to criteria and evaluate the data sources. However, there were some discussions on the exact definitions of requirements sources and criteria. For example, the product vision could have no lifetime value at all, if it is considered as a document. However, the vision could have the highest possible lifetime value as realizing the vision is the company's purpose.

\textit{Interpretation of the results:}

\begin{figure}[!ht]
\centering
\includegraphics[width=0.5\textwidth]{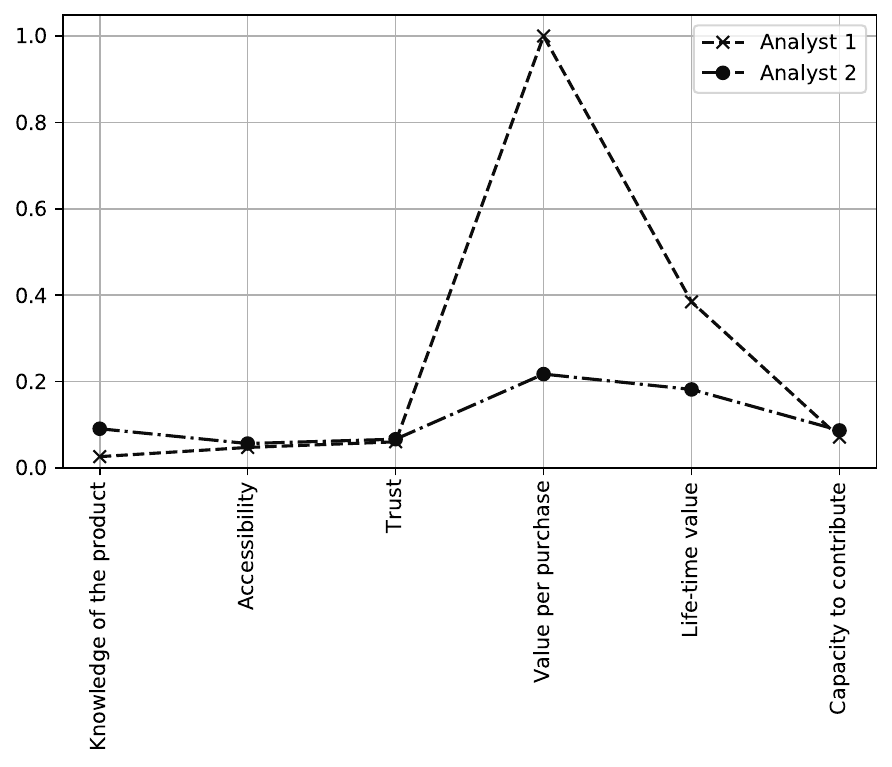}
\caption{Discrepancies in analyst estimates evaluating the relevance of new customers from Case I. Y-axis denotes normalized estimates on how much data from new customers contribute to each criterion. Observe the disagreement between analysts on how much value per purchase new customers deliver.}
\label{fig_results_1a}
\end{figure}

We observe several tendencies when looking at the results; see also Fig.~\ref{fig_case1beforeafter}.
First, the method results show mostly negligible differences between analysts views on how relevant each source 
is. The moderate disagreement concerns the relevance of new customers as a source of data. 
Examining this discrepancy, we found that the root cause is a disagreement on how much value per purchase new customers deliver (see Fig.~\ref{fig_results_1a}). Analyst 1 wished to emphasize that new customers directly contribute to the bottom line of the company. However, the other analyst expressed a view that new customers contribute relatively little compared to established customers.

Secondly, comparing the original estimates and results from the method, see Fig.~\ref{fig_case1beforeafter}, we observe that the method arrives substantially more consistent results between both analysts. Such results demonstrate that the method is useful for consensus-building among the analysts. It also enables to conduct a multi-layer analysis of the analysts' assessments. In that way, it further facilitates the interpretation and better understanding of the selected criteria, the used data sources and the connections between them.

\textit{Lessons learned:} When presented with the aims of our method, practitioners reflected on the challenging nature of selecting requirements sources. Thus, their feedback added support to our hypothesis on the need for a structured method. The participants were already familiar with all the concepts and their relationships. Thus we did not identify any shortcomings in the construct validity. 

The discussions revealed that the main area for improvement could be more refined guidelines for formulating criteria applicable to human stakeholders and other types of requirements sources.

The application of the method helped participants to reflect on their views on what are the most relevant sources of input and why. Specifically, how the use of certain requirements sources connects to short and long-term revenue.

According to the participants, the most valuable outcome from the demonstration was the discovery that even though they consider product vision the key input into product decisions, they lack a clear connection between product vision, revenue, and features. Features that originate from product vision lack an appraisal of potential impact revenue. However, features requested by specific customers can be directly tied to revenue. As a consequence, they are not able to weigh their generalized vision against the needs of a specific customer, thus lacking support for prioritizing requirements.

\subsection{Case II - Construction Equipment}\label{sec_case_2}

The second case is a company developing and manufacturing construction and mining machinery. The machines are software-intensive, and most new features concern updates in the software. The company uses its global partner network to sell and service its machines, and employs over 13 000 people.

Two managers participated in the workshop. The managers are involved and oversee the research and development of new product features and have been with the organization for several years. We followed the steps described in Section~\ref{sec_step4} and started by presenting the method, and then collected the data using a spreadsheet. We did follow-up interviews with each participant to discuss and interpret the results from the method. 

\textit{Reflections on the challenge:}
The participants noted that the organization recognises the challenge and had developed an internal framework for gathering information concerning future product development. However, the application of the framework varies per use. 

\textit{Example case:} The company introduced connectivity to construction machines. The connectivity provides telemetry on machine performance, error reports, maintenance needs, and operator behavior, among other data.

\textit{Triggers:} The opportunity to benefit from the recent IoT development is considered a trigger. 

\textit{Analysts:} The company has a dedicated product planning group for product development. They also involve key accounts and dealers in product decisions. 

\textit{Problem statement:}
A substantial part of the company's offering is the maintenance of the machines to ensure efficient operation and reduce downtime. However, there is limited feedback to product development why a machine breaks down and what usage patterns led to it. IoT technology delivers real-time data on how machines are operated, thus enabling the company to prevent breakdowns and proactively improve the machines. This reduces maintenance costs and improves availability, leading to improved service quality and reduced operational costs.

\textit{Criteria:} The participants needed guidance for selecting criteria, thus they adopted the high-level criteria from Table~\ref{table_criteria}.
The selected criteria were knowledge, amount of experience (with the product), revenue, cost, opportunity cost, timing, ease of use, and suitability for collaboration. Only two analysts participated in the workshop; thus, shortlisting of criteria was not necessary. 





\begin{figure*}[h]
      \centering
      \includegraphics[width=\textwidth]{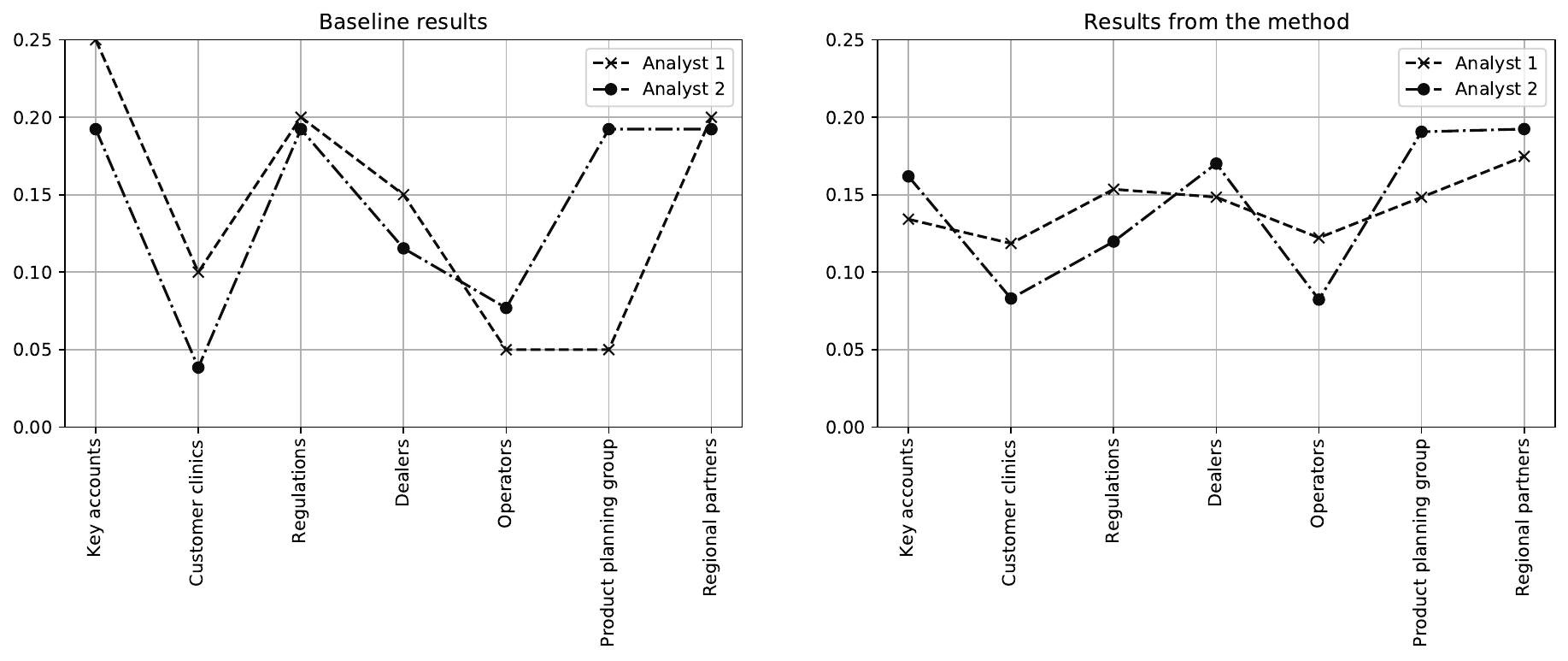}
      \caption{Results from the Case II. 
      Figure on the left shows baseline estimates, i.e. respondent estimates without the method. Figure on the right shows results from applying the method. 
      Y-axis denotes the relative importance of data sources. 
      }
      \label{fig_case2beforeafter}
\end{figure*}

\textit{Requirements sources:} The participants reflected that the key sources for ideas are dealers and key accounts. Dealers and regional partners work closely with end-customers and have first-hand knowledge of their actual needs. Furthermore, key accounts and dealers provide substantial revenue for the company; thus there is a financial incentive to prioritize their input. Other sources of information are industry standards, regulations, engineering, direct feedback from customers, workshops with machine operators, and engineers. We show the initial baseline ranking of data sources in Fig.~\ref{fig_case2beforeafter}, the left side.

\textit{Application of the method:}
The method application ran with only minor issues. The first issue was identifying a problem to be studied, but after the initial discussion, the participants quickly agreed. The effort needed to get them acquainted was moderate. Most time was dedicated towards the individual estimates of the importance of each requirements source against the criteria (about 45 minutes on average for each participant). Some additional effort was needed during the prioritization to explain the meaning of particular criteria for a particular data source.  

\textit{Interpretation of the results:}


The baseline estimates  (Fig.~\ref{fig_case2beforeafter}, left side) show a discrepancy between the analysts regarding the product planning group's importance. However, applying the method (Fig.~\ref{fig_case2beforeafter}, right side) helped to arrive at more aligned results. The key accounts were estimated (without the method) to be the primary data source. However, the method suggests that input from regional partners, dealers, and product planning groups should be prioritized.






\textit{Lessons learned:}
The participants' reflections during the demonstration supported our earlier hypothesis that product decisions are often opinion-based. The application of the method builds consensus and helps to reduce discrepancies.

We observed a need to accommodate a scenario where distributed analysts work independently, and combine their input. This can be achieved by more stringent guidelines on selecting criteria and data sources for evaluation and tool support to facilitate the method. Our participants saw value in eliciting the criteria and consensus building among them, and a large group of stakeholders involved in decision-making.  

\subsection{Case III - Real Estate Management}


PlanOn Software offers solutions for real estate managers, occupiers, and service providers. The main offering of the company is a software platform streamlining real estate management. The main office of PlanOn is located in the Netherlands. However, the company services customers in 40 countries. The company has operated since 1982 and now employs over 750 people. 

We conducted three workshops with this company. The first workshop was to explore their case, the second to apply the method and collect data, and the third workshop to analyze the results. A total of six managers from different areas participated in the workshops. The participants were:

\begin{itemize}
  \item Sales manager
  \item Key account manager
  \item Systems design/development manager
  \item Functional design/product manager
  \item Marketing manager
  \item Lead product manager
\end{itemize}

The participants have varying experiences in the organization ranging from 1 to 10 years.

\textit{Reflections on the challenge:} The participants immediately pointed out that most product decisions are heavily influenced by one of the founders (currently an executive in the company), who maintains great influence on the organization and offers extensive experience and knowledge of the market.  

The participants shared some concerns about the status quo and indicated that product decision-making needs to be distributed to sustain further growth of the organization.

Marketing and sales managers mentioned that they are typically involved late in product development, leaving them with impaired influence on marketing and sales activities. 

In addition to internal sources, the participants mentioned piloting new features with focus groups and useful contacts with key customers.  

\textit{Triggers:} The company decided to improve its offering by providing a cost settlement feature to mediate relationships between utility service providers and tenants. 
At the time of our workshops, the feature was recently launched.

\textit{Analysts: } There is no established way of product decision-making. Most decisions are made by the before-mentioned executive. Six internal stakeholders were invited into the workshop.

\textit{Problem statement:} Since there is no established process of selecting data sources, participants wished to explore the rationale behind each source. Most importantly, to what extent the executive's knowledge and experience should be prioritized over market signals.

\textit{Requirements sources:} During the discussions, we elicited several potential requirements sources they are currently using or would consider useful. The participants identified market research, marketing, prospects (potential customers), key customers panel, sales, external analysts, the executive, launch customers, product maintenance teams, product telemetry, engineering teams, standards/regulations, competitors, and accounting principles as potential data sources.

\begin{figure*}[ht]
      \centering
      \includegraphics[width=\textwidth]{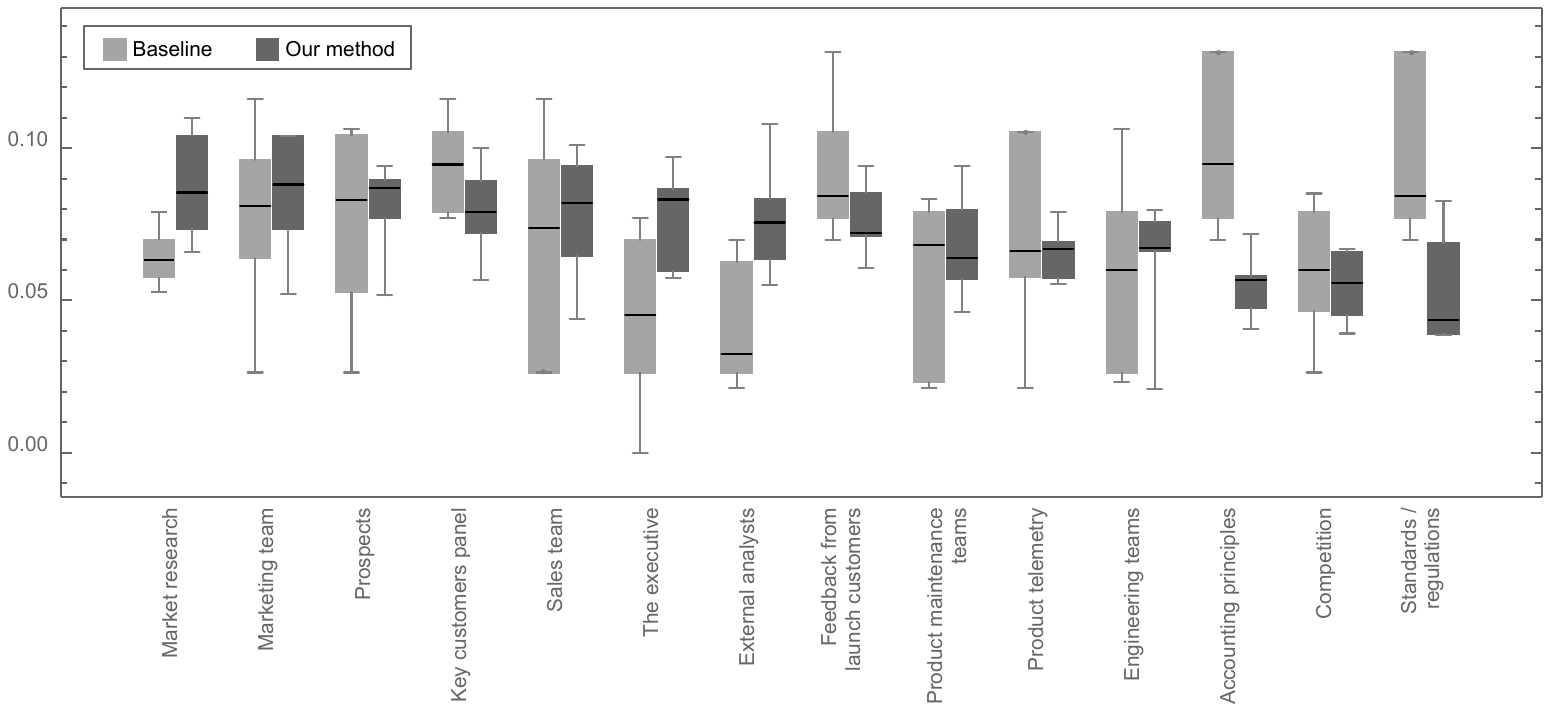}
      \caption{Results from the Case III, comparing ad-hoc estimates (baseline) and results of the method on the importance of data sources.
      Y-axis denotes the relative importance of data sources. 
      }
      \label{fig_case3beforeafter}
\end{figure*}

Before introducing the method, we asked participants to evaluate the data sources according to their relevance to the cost settlement feature. We asked each participant to evaluate each source on a scale 0 - 5. The results are summarized in Fig.~\ref{fig_case3beforeafter}. Note that we visualize analyst estimates with box plots to improve the readability of a large number of data points.

\textit{Criteria: } The participants identified knowledge about market direction (from market leaders), knowledge about the mainstream market's needs, potential monetary gains, investment in accessing a source, knowledge about technical possibilities and constraints, risk of tunnel vision, and risk of non-compliance as relevant criteria. The participants discussed the criteria and swiftly reached a consensus on what criteria were relevant. Thus, explicit shortlisting of criteria was not performed.

\begin{figure*}[ht]
      \centering
      \includegraphics[width=\textwidth]{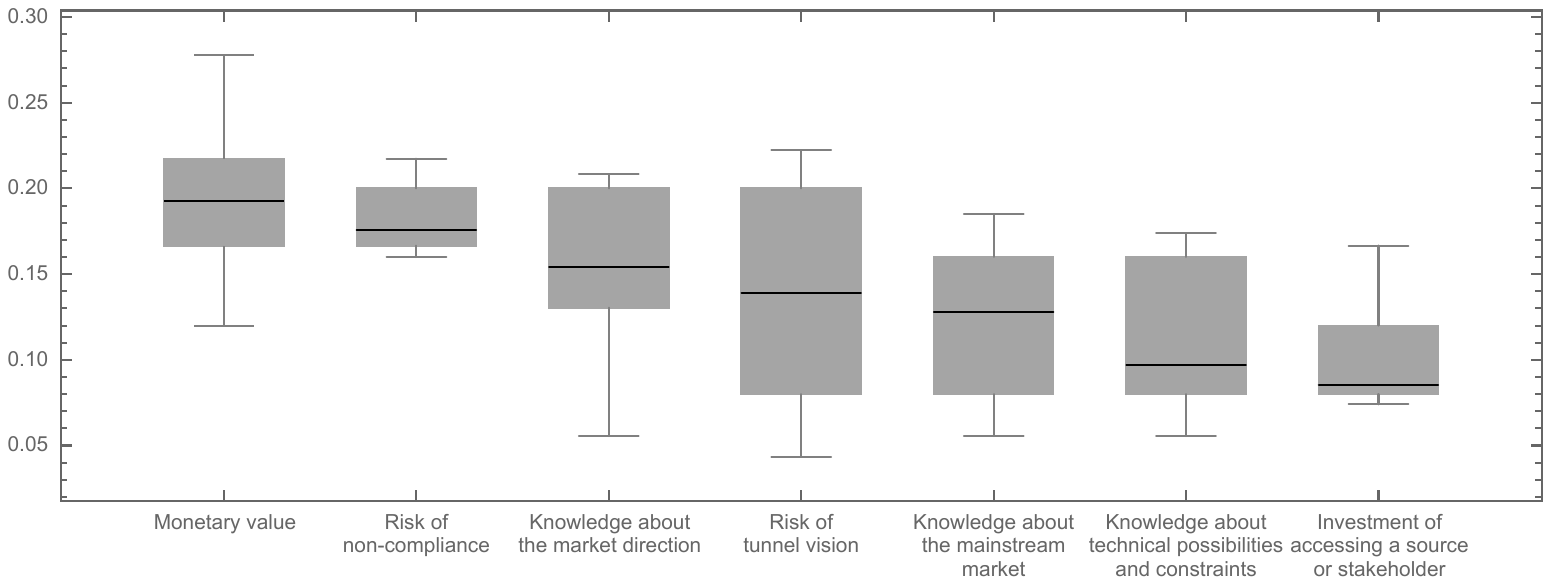}
      \caption{Results from the Case III, showing the participant estimates on criteria importance.
      Y-axis denotes the normalized relative importance of data sources. 
      }
      \label{fig_case3criteria}
\end{figure*}

We show the summary of participant estimates on criteria importance in Fig.~\ref{fig_case3criteria}.

\textit{Application of the method:} The application of the method went smoothly despite the larger number of participants. The steps of our method were easily understood and followed.  

We parallelized the data collection by distributing individual spreadsheets to participants; thus, the larger number of participants had a negligible effect on the time spent.

\textit{Interpretation of the results: }
Analyzing the results, see Figure ~\ref{fig_case3beforeafter}, we observe several tendencies. After applying the method, analysts' estimates are less spread, indicating better agreement. Discussing the results, participants admitted that the use of criteria helped them focus and generate insight into the rationale behind using each data source.

We observe that standards, regulations, and accounting principles have lost some importance according to the method. Analysts explained that to ensure global relevance, they specifically avoid building features that depend on local laws. Thus, as long as they are not violating any regulations, they do not consider them as relevant input to product development.

Interestingly, the executive has gained some importance as a requirements source. We investigated this further and found that the executive is evaluated high on criteria such as the vision of further market direction and mitigating the risk of tunnel vision.

When we presented these results to the participants, they admitted the executive is the main source of product vision, due to long experience in the domain. Furthermore, the executive is the only person overseeing the work of different teams and mitigating silo effects that may occur.

We further looked at what other data sources could be used to replace knowledge provided by the executive and found that external market analysis could be used as input for determining product vision and further market direction. 

\textit{Lessons learned:} From Case III we have learned that the challenge of selecting requirements sources is relevant. PlanOn realizes the pitfalls of following the vision of a single executive or customer, however lacks tools for systematically analyzing requirements sources.

Our method scales well with an increased number of analysts. The application of the method, including discussions, elicitation of evaluations, and analysis fit in a two hours slot. However, an on-line tool facilitating the method could streamline the application and remove the need for manual handling of spreadsheets.

In conclusion, participants reflected that the method had helped them to  understand their requirements sources and had provoked some interesting discussions.

\subsection{Potential improvements and limitations}

One of the central areas for improvement is providing better support for identifying relevant criteria and candidate requirements sources. Using overly broad criteria and  requirements sources reduces the usefulness of the results. However, identification and evaluations of many very specific terms may require excessive effort. To alleviate this shortcoming, we propose to iteratively apply the method by increasing detail with each iteration. We suggest starting with high-level requirements sources and criteria, e.g., ones given in Tables~\ref{table_criteria} and \ref{table_sources}. Then gradually eliminate less relevant criteria and detail the relevant ones, to arrive at specific data instances (e.g., Customer 1 and Customer 2) from high level classes  (e.g., key customers). 

Interestingly, in all three cases analysts were able to agree on a list of relevant criteria without any issues. That is, there were no extreme opinions on whether a criterion is relevant or not. Thus, we did not perform the shortlisting of criteria. Explicit shortlisting of criteria, e.g., by voting, is relevant only if an overly large number of criteria are identified, and there is a strong disagreement between analysts on what criteria are worth considering further.




We use an ordinal scale from 0 to 5 to facilitate analysts in the task of ranking the criteria~\cite{labovitz1970assignment}. Alternative ranking methods might also be used, for example, ones based on pairwise comparison~\cite{barzilai1997deriving} and large preference relation~\cite{Fodor1998}. However, evaluations in the ordinal scale are more intuitive, faster, and require relatively fewer steps than the other two methods.

Another limitation is the use of arithmetic mean, which is well-known to be sensitive to extreme values. In general, the choice of the aggregation operator is critical since some aggregation operators can lead to a significant loss of information as extreme values  can greatly influence their scores (e.g., the arithmetic mean), while others are penalizing too much for low-scoring outliers (e.g., the geometric mean and the harmonic mean)~\cite{bullen2013means}. A possible solution to the described problem is to use different aggregation operators in order to find some trade-off between their conflicting behavior, e.g., hybrid aggregation operators proposed in Tsiporkova et al.~\cite{Tsiporkova2004,Tsiporkova2006}.

A potential scalability issue emerges when inputs from many analysts, a large number of criteria, or data sources need to be considered. A large number of criteria ($C$) and data sources ($D$) require each analyst to provide  $C + D \times C$ inputs. However, this limitation remains theoretical until full-scale validation of the method.

\section{Discussion}\label{sec_discussion}
We structure our discussion around the research questions and objectives of the method presented in Section~\ref{sec_objectives}. We also present lessons learned and improvement suggestions.

\subsection{What are the needs towards identifying and selecting the most relevant requirements sources (RQ1 and RQ2)}

\textbf{Objective 1: Context independence of the method.} The review of related work in Section~\ref{sec_rem} reveals that stakeholders are usually identified once at the beginning of the project and based on generalized characteristics~\cite{babar2014stakeholder,babar2015stakemeter,ballejos2011modeling}. Such an approach does not take into consideration the dynamic nature of software-intensive  product development with a continuous inflow of requirements~\cite{Karlsson2007} and many changes to the initial scope of the project~\cite{Wnuk2015TSE}. Therefore, we formulated in Objective 1 that the method should support the adaptation to new or changed problems or tasks. The results from the case studies indicate that Objective 1 has been achieved. Our method applied to the supply chain digitalization case helped decision-makers in understanding the un-articulated needs (lagging sellers) and the need to amplify this source of requirements. We also observed that our method helps to evaluate the importance of current customers vs. potential new customers. Our results confirm that changing business environment needs to be reflected in adaptation of requirements sources and their ranks, increasing the flexibility in responding to changing needs, and new business objectives. Our results support the assumption that the selection of data sources needs to be specific to each situation (\textit{Objective 1}). This reflects the current trend in decision making where information networks get created for a specific decision problem and are dissolved after a decision is made~\cite{van2014new}. 

Our proposed method is not coupled with any specific context, criteria, or data sources. Instead, we provide instructions on how to identify inputs for the method from its context of use. Thus, the method can be tailored for use in a wide range of situations, that we exemplify providing example criteria and requirements sources.

\textbf{Objectives 2 and 3: Support for identification and ranking of different sources.} Given the dynamic nature of software business and the advent of data-driven product strategies, the identification of new and unrealized requirements sources becomes critical (\textit{Objective 2}). Our method fulfills this objective thanks to inviting multiple roles and considering multiple criteria associated with the problem or task under consideration.  

In a crowd requirements engineering context, analysts may benefit from a wide range of requirements sources, both within companies' control and beyond~\cite{groen2017crowd,alexander2005taxonomy}. Thus, directly consulting specific stakeholder groups may not be possible or require substantial efforts. In turn, stakeholder needs could be elicited indirectly, for example, by examining documents, artifacts, and data generated by stakeholders interacting with the product. Our method can support the selection of a broad range of requirements  in such a context (\textit{Objective 3}).

There are no inherent limitations on what types of requirements sources the method can operate with. One difficulty could be to formulate evaluation criteria that suit both individuals and potential requirements sources. Thus, the formulation of criteria and requirements sources can be done iteratively. The initial set of criteria can help understand the needs for requirements sources, thus supporting the identification of new requirements sources. 

\textbf{Objective 4: Consensus building and collaboration between multiple roles}
Requirements engineering is a collaborative activity, and product decisions could be cross-cutting and require consideration of multiple perspectives~\cite{Gorschek2006a, Barczak2010}. Aggregating multiple perspectives and reaching consensus between opposing views is challenging. Our method helps discover and resolve inconsistencies between the roles around the importance of the data sources for particular criteria. 

Handling inconsistencies during requirements prioritization has been a significant challenge for many prioritization techniques. For example, AHP includes consistency checks for pairwise comparison~\cite{karlsson1997cost}. Our proposed method pro-actively reduces conflicts between stakeholders. The results from all three cases show a flattening of the degree of disagreement between the participants. Our work also complements previous efforts on studying prioritization imbalances, e.g., favoring fewer large customers over many  smaller customers~\cite{Taylor2011PrioritizationIBM}.

Our method proposes to collect individual preferences in multiple steps and at each step, analyze the discrepancies between the preferences. In this way, we mitigate the adverse effects of group-based approaches and encourage focused discussions on specific disagreements.

\textbf{Objective 5: Transparency and understandability.} Visualizing results helps to identify discrepancies and increase transparency and understandability of the results (\textit{Objective 5}). 
In the third case, product decisions are dominated by a single executive. Although his knowledge and expertise is valuable, the organization must move towards replicating his input. Our method helped the organization to identify the strengths of the executive. By analyzing other data sources, it was possible to identify alternative data sources to remove the dependency on a single stakeholder in critical decisions.

\textbf{Effort and value return.} An important concern is to what extent the benefits of using a method outweigh the resources needed to use it. Using our method requires several analysts to meet, set forth relevant criteria and candidate data sources, run the method, and discuss the results. We argue that such meetings already take place, and the use of the method adds structure and transparency to these meetings. Moreover, decision on criteria helps to bring common understanding about the aspects used to select and prioritize requirements sources.  

In two cases, participants have regular product steering group meetings. Thus, using the method to support such meetings would require minimal additional effort. Not utilizing any structured approach could lead to unproductive discussions and poor decisions. At a minimum, a poor decision requires an immediate extra effort to correct it. However, it could also lead to wasted development effort and missed market opportunities. Therefore, potential benefits from using the method outweigh the additional effort.



\subsection{RQ3: Suggested method improvements}

We summarize received feedback and our lessons learned from the workshops. In all three workshops, we identified the need for additional guidance for identifying and interpreting the evaluation criteria. Even though we exemplify some criteria in the method description, the selection of criteria was difficult. The support for criteria selection could be provided with, for example, a taxonomy of criteria and their descriptions. 

In all three companies, the practitioners were able to list relevant requirements sources based on their experience and our examples. These were mostly current or well known requirements sources. However, the identification of candidate/future requirements sources could be supported further by a more extensive taxonomy and improved guidelines.

We implemented the method using spreadsheets and a simple script. During the application of the method, participants were guided and supported by researchers. However, we identify the need for a robust tool guiding analysts through the steps of the method and minimizing the need for guidance from researchers. This tool should have examples and suggestions how to proceed in the process.

\section{Conclusions and further work}\label{ch8:sec_conclusions}
In this paper, we have proposed and demonstrated a method for the selection of requirements sources. The method comprises of systematic steps to collaboratively identify candidate requirements sources, evaluate them according to agreed criteria, and to produce a ranking of their relevance to the decision. We have paid particular attention to consensus building and provided means to analyze and understand the final results.

We have explored the usability of the method in three different industrial scenarios and have gathered data for its further refinement. The method has been shown to facilitate consensus building among analysts and additionally contribute to a better understanding and a consistent view of the importance of individual requirements sources. The systematic steps of the method enable analysts to analyze and remedy their disagreements.

An important insight from the cases is that practitioners use ad-hoc methods based on accessibility and convenience in requirements sources selection. Furthermore, analysts views on which sources to prioritize vary significantly. After piloting our method, practitioners reported that the use of the structured method allowed them to reflect and arrive at a joint view on the importance of different data sources.

The case study results show that our method can improve the selection of requirements sources by highlighting and helping to resolve discrepancies in the analyst's views. The method also mitigates the issue of one list of requirements sources composed by one person and kept mostly unchanged over time. The workshops with the three companies have highlighted that understanding the rationale and value behind already known requirements sources is a strength of our method, and exploring other views on potential new data sources is beneficial for our company partners. Furthermore, our method can help to discover new data sources that are currently outside of the organization's perspective~\cite{kolpondinos2020garuso}. As our method is triggered by a specific event or program, its nature becomes iterative, and it is re-run on a regular basis. This, in turn, enables updates of potential data sources and adaptations towards new product opportunities and problems.

Further work focuses on additional validations and improvement rounds. We plan to  develop a tool to facilitate the use of the method and data collection on its application, and extend our support towards criteria and data sources identification.

\section*{Acknowledgements}

The authors would like to thank participants of the demonstrations for their time dedicated to this study.  

We would like to acknowledge that this work was supported by the(removed for double blind review)
KKS foundation through the S.E.R.T. Research Profile project at Blekinge Institute of Technology.





\bibliographystyle{elsarticle-num}
\bibliography{library.bib}

\begin{thebibliography}{10}
\expandafter\ifx\csname url\endcsname\relax
  \def\url#1{\texttt{#1}}\fi
\expandafter\ifx\csname urlprefix\endcsname\relax\def\urlprefix{URL }\fi
\expandafter\ifx\csname href\endcsname\relax
  \def\href#1#2{#2} \def\path#1{#1}\fi

\bibitem{hofmann2001requirements}
H.~F. Hofmann, F.~Lehner, Requirements engineering as a success factor in
  software projects, IEEE software~(4) (2001) 58--66.

\bibitem{SwebookV3}
{IEEE Computer Society}, Guide to the Software Engineering Body of Knowledge
  (SWEBOK(R)): Version 3.0, 3rd Edition, IEEE Computer Society Press, Los
  Alamitos, CA, USA, 2014.

\bibitem{alexander2005taxonomy}
I.~F. Alexander, A taxonomy of stakeholders: Human roles in system development,
  International Journal of Technology and Human Interaction (IJTHI) 1~(1)
  (2005) 23--59.

\bibitem{hamka2014mobile}
F.~Hamka, H.~Bouwman, M.~De~Reuver, M.~Kroesen, Mobile customer segmentation
  based on smartphone measurement, Telematics and Informatics 31~(2) (2014)
  220--227.

\bibitem{anwar2016stakeholders}
F.~Anwar, R.~Razali, Stakeholders selection model for software requirements
  elicitation, American Journal of Applied Sciences 13~(6) (2016) 726--738.

\bibitem{Regnell2005}
B.~Regnell, S.~Brinkkemper, Market-Driven Requirements Engineering for Software
  Products, Springer Berlin Heidelberg, Berlin, Heidelberg, 2005, pp. 287--308.

\bibitem{maalej2016toward}
W.~Maalej, M.~Nayebi, T.~Johann, G.~Ruhe, Toward data-driven requirements
  engineering, IEEE Software 33~(1) (2016) 48--54.

\bibitem{barik2016bones}
T.~Barik, R.~DeLine, S.~Drucker, D.~Fisher, The bones of the system: A case
  study of logging and telemetry at microsoft, in: 2016 IEEE/ACM 38th
  International Conference on Software Engineering Companion (ICSE-C), IEEE,
  2016, pp. 92--101.

\bibitem{ProductCMMIfor2006}
{Software Engineering Institute}, {CMMI for Development, Version 1.2}, Tech.
  rep., Software Engineering Institute, Carnegie Mellon University (2006).

\bibitem{KruchtenRUP}
P.~Kruchten, The Rational Unified Process: An Introduction, Second Edition, 2nd
  Edition, Addison-Wesley Longman Publishing Co., Inc., Boston, MA, USA, 2000.

\bibitem{ISO12207}
R.~Singh, International standard iso/iec 12207 software life cycle processes,
  Software Process: Improvement and Practice 2~(1) (1996) 35--50.

\bibitem{Pressman2001}
R.~S. Pressman, Software Engineering: A Practitioner's Approach, 5th Edition,
  McGraw-Hill Higher Education, 2001.

\bibitem{Sommerville2010}
I.~Sommerville, Software Engineering, 9th Edition, Addison-Wesley Publishing
  Company, USA, 2010.

\bibitem{Lauesen2001}
S.~Lauesen, Software Requirements: Styles and Techniques, 1st Edition, Pearson
  Education, 2001.

\bibitem{Sharp1999}
H.~Sharp, A.~Finkelsteiin, G.~Galal, Stakeholder identification in the
  requirements engineering process, in: 10th International Workshop on Database
  \& Expert Systems Applications, DEXA '99, IEEE Computer Society, Washington,
  DC, USA, 1999, pp. 387--.

\bibitem{Preiss2001}
O.~Preiss, A.~Wegmann, Stakeholder discovery and classification based on
  systems science principles, in: Proceedings of the Second Asia-Pacific
  Conference on Quality Software, APAQS '01, IEEE Computer Society, Washington,
  DC, USA, 2001, pp. 194--.

\bibitem{von2006democratizing}
E.~Von~Hippel, Democratizing innovation, the MIT Press, 2006.

\bibitem{Dahlstedt}
{\AA}.~G. Dahlstedt, L.~Karlsson, A.~Persson, J.~{Natt och Dag}, B.~Regnell,
  {Market-Driven Requirements Engineering Processes for Software Products - a
  Report on Current Practices}, in: International Workshop on COTS and Product
  Software, RECOTS 2003, 2003.

\bibitem{Klotinsc}
E.~Klotins, M.~Unterkalmsteiner, T.~Gorschek, {Software Engineering
  Anti-patterns in start-ups}, In review by IEEE Software.

\bibitem{Pacheco2012}
C.~Pacheco, I.~Garcia, {A systematic literature review of stakeholder
  identification methods in requirements elicitation}, Journal of Systems and
  Software 85~(9) (2012) 2171--2181.

\bibitem{Karlsson2007}
L.~Karlsson, {\AA}.~G. Dahlstedt, B.~Regnell, J.~N. och Dag, A.~Persson,
  Requirements engineering challenges in market-driven software development--an
  interview study with practitioners, Information and Software technology
  49~(6) (2007) 588--604.

\bibitem{Klotins2016}
E.~Klotins, M.~Unterkalmsteiner, T.~Gorschek, Software engineering in start-up
  companies: An analysis of 88 experience reports, Empirical Software
  Engineering 24~(1) (2019) 68--102.

\bibitem{Alves2006}
C.~Alves, S.~Pereira, J.~Castro, {A study in market-driven requirements
  engineering}, Workshop em Engenharia de Requisitos WER06 (2006) 2--3.

\bibitem{Liu}
{Xiaoqing Liu}, C.~S. {Veera}, Y.~{Sun}, K.~{Noguchi}, Y.~{Kyoya}, Priority
  assessment of software requirements from multiple perspectives, in:
  Proceedings of the 28th Annual International Computer Software and
  Applications Conference, 2004. COMPSAC 2004., 2004, pp. 410--415 vol.1.
\newblock \href {http://dx.doi.org/10.1109/CMPSAC.2004.1342872}
  {\path{doi:10.1109/CMPSAC.2004.1342872}}.

\bibitem{saaty1988analytic}
T.~L. Saaty, What is the analytic hierarchy process?, in: Mathematical models
  for decision support, Springer, 1988, pp. 109--121.

\bibitem{babar2015stakemeter}
M.~I. Babar, M.~Ghazali, D.~N. Jawawi, K.~B. Zaheer, Stakemeter: Value-based
  stakeholder identification and quantification framework for value-based
  software systems, PloS one 10~(3).

\bibitem{burnay2016stakeholders}
C.~Burnay, Are stakeholders the only source of information for requirements
  engineers? toward a taxonomy of elicitation information sources, ACM
  Transactions on Management Information Systems (TMIS) 7~(3) (2016) 8.

\bibitem{groen2017crowd}
E.~C. Groen, N.~Seyff, R.~Ali, F.~Dalpiaz, J.~Doerr, E.~Guzman, M.~Hosseini,
  J.~Marco, M.~Oriol, A.~Perini, et~al., The crowd in requirements engineering:
  The landscape and challenges, IEEE software 34~(2) (2017) 44--52.

\bibitem{jin2016understanding}
J.~Jin, Y.~Liu, P.~Ji, H.~Liu, Understanding big consumer opinion data for
  market-driven product design, International Journal of Production Research
  54~(10) (2016) 3019--3041.

\bibitem{alspaugh2013ongoing}
T.~A. Alspaugh, W.~Scacchi, Ongoing software development without classical
  requirements, in: Requirements Engineering Conference (RE), 2013 21st IEEE
  International, IEEE, 2013, pp. 165--174.

\bibitem{klotins2019progression}
E.~Klotins, M.~Unterkalmsteiner, P.~Chatzipetrou, T.~Gorschek, R.~Prikladniki,
  N.~Tripathi, L.~Pompermaier, A progression model of software engineering
  goals, challenges, and practices in start-ups, IEEE Transactions on Software
  Engineering.

\bibitem{razali2011selecting}
R.~Razali, F.~Anwar, Selecting the right stakeholders for requirements
  elicitation: a systematic approach, Journal of Theoretical and Applied
  Information Technology 33~(2) (2011) 250--257.

\bibitem{hujainah2018stakeholder}
F.~Hujainah, R.~B.~A. Baka, B.~Al-Haimi, M.~A. Abdulgabber, Stakeholder
  quantification and prioritisation research: A systematic literature review,
  Information and Software Technology.

\bibitem{babar2014stakeholder}
M.~I. Babar, M.~Ghazali, D.~N. Jawawi, A.~Elsafi, Stakeholder management in
  value-based software development: systematic review, IET Software 8~(5)
  (2014) 219--231.

\bibitem{gu2011taxonomy}
Q.~Gu, M.~Parkin, P.~Lago, A taxonomy of service engineering stakeholder types,
  in: European Conference on a Service-Based Internet, Springer, 2011, pp.
  206--219.

\bibitem{bano2014systematic}
M.~Bano, D.~Zowghi, N.~Ikram, Systematic reviews in requirements engineering: A
  tertiary study, in: 2014 IEEE 4th International Workshop on Empirical
  Requirements Engineering (EmpiRE), IEEE, 2014, pp. 9--16.

\bibitem{aurum2003fundamental}
A.~Aurum, C.~Wohlin, The fundamental nature of requirements engineering
  activities as a decision-making process, Information and Software Technology
  45~(14) (2003) 945--954.

\bibitem{Petersen2009a}
K.~Petersen, C.~Wohlin, {Context in industiral software engineering research},
  {\ldots} Symposium on Empirical Software Engineering {\ldots}.

\bibitem{Ivarsson2010}
M.~Ivarsson, T.~Gorschek, {A method for evaluating rigor and industrial
  relevance of technology evaluations}, Empirical Software Engineering 16~(3)
  (2010) 365--395.

\bibitem{babar2014bi}
M.~I. Babar, M.~Ghazali, D.~N. Jawawi, A bi-metric and fuzzy c-means based
  intelligent stakeholder quantification system for value-based software., in:
  SoMeT, 2014, pp. 295--309.

\bibitem{bendjenna2012using}
H.~Bendjenna, P.-J. Charre, N.~Eddine~Zarour, Using multi-criteria analysis to
  prioritize stakeholders, Journal of Systems and Information Technology 14~(3)
  (2012) 264--280.

\bibitem{mcmanus2004stakeholder}
J.~McManus, A stakeholder perspective within software engineering projects, in:
  2004 IEEE International Engineering Management Conference (IEEE Cat. No.
  04CH37574), Vol.~2, IEEE, 2004, pp. 880--884.

\bibitem{ballejos2011modeling}
L.~C. Ballejos, J.~M. Montagna, Modeling stakeholders for information systems
  design processes, Requirements engineering 16~(4) (2011) 281--296.

\bibitem{lim2010stakenet}
S.~L. Lim, D.~Quercia, A.~Finkelstein, Stakenet: using social networks to
  analyse the stakeholders of large-scale software projects, in: Proceedings of
  the 32Nd ACM/IEEE International Conference on Software Engineering-Volume 1,
  ACM, 2010, pp. 295--304.

\bibitem{whitehead2007collaboration}
J.~Whitehead, Collaboration in software engineering: A roadmap, in: Future of
  Software Engineering (FOSE'07), IEEE, 2007, pp. 214--225.

\bibitem{matsatsinis2001mcda}
N.~F. Matsatsinis, A.~P. Samaras, Mcda and preference disaggregation in group
  decision support systems, European Journal of Operational Research 130~(2)
  (2001) 414--429.

\bibitem{tindale2000social}
R.~S. Tindale, T.~Kameda, ‘social sharedness’ as a unifying theme for
  information processing in groups, Group Processes \& Intergroup Relations
  3~(2) (2000) 123--140.

\bibitem{Broniewicz}
E.~Broniewicz, K.~Ogrodnik, \href{https://www.mdpi.com/1996-1073/14/16/5100}{A
  comparative evaluation of multi-criteria analysis methods for sustainable
  transport}, Energies 14~(16).
\newblock \href {http://dx.doi.org/10.3390/en14165100}
  {\path{doi:10.3390/en14165100}}.
\newline\urlprefix\url{https://www.mdpi.com/1996-1073/14/16/5100}

\bibitem{Triantaphyllou}
E.~Triantaphyllou, K.~Baig, The impact of aggregating benefit and cost criteria
  in four mcda methods, IEEE Transactions on Engineering Management 52~(2)
  (2005) 213--226.
\newblock \href {http://dx.doi.org/10.1109/TEM.2005.845221}
  {\path{doi:10.1109/TEM.2005.845221}}.

\bibitem{5353151}
U.~e~Habiba, S.~Asghar, A survey on multi-criteria decision making approaches,
  in: 2009 International Conference on Emerging Technologies, 2009, pp.
  321--325.
\newblock \href {http://dx.doi.org/10.1109/ICET.2009.5353151}
  {\path{doi:10.1109/ICET.2009.5353151}}.

\bibitem{wieringa2014design}
R.~J. Wieringa, Design science methodology for information systems and software
  engineering, Springer, 2014.

\bibitem{tovar2006stakeholder}
E.~Tovar, C.~Pacheco, Stakeholder identification in requirements engineering:
  Comparison of methods, in: Proc. of Software Engineering Applications (SEA),
  2006.

\bibitem{Gorschek2006}
T.~Gorschek, C.~Wohlin, P.~Garre, S.~Larrson, {A model for technology transfer
  in practice}, Learning and Leading with Technology 30~(4) (2006) 88--95.

\bibitem{Runeson2012}
P.~Runeson, M.~H{\"{o}}st, A.~Rainer, B.~Regnell, {Case study research in
  software engineering}, John Wiley {\&} Sons, Inc., 2012.

\bibitem{Flyvbjerg2006}
B.~Flyvbjerg, {Five Misunderstandings About Case-Study Research}, Qualitative
  Inquiry 12~(2) (2006) 219--245.

\bibitem{annamalai2013importance}
N.~Annamalai, S.~Kamaruddin, I.~Abdul~Azid, T.~Yeoh, Importance of problem
  statement in solving industry problems, in: Applied Mechanics and Materials,
  Vol. 421, Trans Tech Publ, 2013, pp. 857--863.

\bibitem{MitchellStakeholderTheory}
R.~K. Mitchell, B.~R. Agle, D.~J. Wood, Toward a theory of stakeholder
  identification and salience: Defining the principle of who and what really
  counts, The Academy of Management Review 22~(4) (1997) 853--886.

\bibitem{riegel2015systematic}
N.~Riegel, J.~Doerr, A systematic literature review of requirements
  prioritization criteria, in: International Working Conference on Requirements
  Engineering: Foundation for Software Quality, Springer, 2015, pp. 300--317.

\bibitem{vigna2015weighted}
S.~Vigna, A weighted correlation index for rankings with ties, in: 24th
  international conference on World Wide Web, 2015, pp. 1166--1176.

\bibitem{labovitz1970assignment}
S.~Labovitz, The assignment of numbers to rank order categories, American
  sociological review (1970) 515--524.

\bibitem{barzilai1997deriving}
J.~Barzilai, Deriving weights from pairwise comparison matrices, Journal of the
  operational research society 48~(12) (1997) 1226--1232.

\bibitem{Fodor1998}
J.~Fodor, S.~Orlovski, P.~Perny, M.~Roubens, The Use of Fuzzy Preference Models
  in Multiple Criteria Choice, Ranking and Sorting, Springer US, Boston, MA,
  1998, pp. 69--101.
\newblock \href {http://dx.doi.org/10.1007/978-1-4615-5645-9_3}
  {\path{doi:10.1007/978-1-4615-5645-9_3}}.

\bibitem{bullen2013means}
P.~S. Bullen, D.~S. Mitrinovic, M.~Vasic, Means and their Inequalities,
  Vol.~31, Springer Science \& Business Media, 2013.

\bibitem{Tsiporkova2004}
E.~Tsiporkova, V.~Boeva, Nonparametric recursive aggregation process,
  Kybernetika, Journal of the Czech Society for Cybernetics and Information
  Sciences 40 (2004) 51--70.

\bibitem{Tsiporkova2006}
E.~Tsiporkova, V.~Boeva, Multi-step ranking of alternatives in a multi-criteria
  and multi-expert decision making environment, Information Sciences 176 (2006)
  2673--2697.

\bibitem{Wnuk2015TSE}
K.~{Wnuk}, T.~{Gorschek}, D.~{Callele}, E.~{Karlsson}, E.~{Åhlin},
  B.~{Regnell}, Supporting scope tracking and visualization for very
  large-scale requirements engineering-utilizing fsc+, decision patterns, and
  atomic decision visualizations, IEEE Transactions on Software Engineering
  42~(1) (2016) 47--74.

\bibitem{van2014new}
A.~Van't~Spijker, The new oil: using innovative business models to turn data
  into profit, Technics Publications, 2014.

\bibitem{Gorschek2006a}
T.~Gorschek, C.~Wohlin, {Requirements abstraction model}, Requirements
  Engineering 11~(1) (2006) 79--101.

\bibitem{Barczak2010}
G.~Barczak, F.~Lassk, J.~Mulki, {Antecedents of team creativity: An examination
  of team emotional intelligence, team trust and collaborative culture},
  Creativity and Innovation Management 19~(4) (2010) 332--345.

\bibitem{karlsson1997cost}
J.~Karlsson, K.~Ryan, A cost-value approach for prioritizing requirements, IEEE
  software 14~(5) (1997) 67--74.

\bibitem{Taylor2011PrioritizationIBM}
C.~A. {Taylor}, A.~V. {Miranskyy}, N.~H. {Madhavji}, Request-implementation
  ratio as an indicator for requirements prioritisation imbalance, in: 2011
  Fifth International Workshop on Software Product Management (IWSPM), 2011,
  pp. 3--6.

\bibitem{kolpondinos2020garuso}
M.~Z. Kolpondinos, M.~Glinz, Garuso: a gamification approach for involving
  stakeholders outside organizational reach in requirements engineering,
  Requirements Engineering 25~(2) (2020) 185--212.

\end{thebibliography}







\end{document}